\documentclass[
  nofootinbib,
  reprint,
  superscriptaddress,
  amsmath,amssymb,
  aps, prx,
  floatfix,
]{revtex4-2}

\usepackage{tabularx}
\usepackage{graphicx}
\usepackage{dcolumn}
\usepackage{bm}
\usepackage{xcolor}
\usepackage{textcomp}
\usepackage{amsmath}
\usepackage{multirow}
\usepackage{tikz}
\usepackage[utf8]{inputenc}
\usepackage{booktabs}
\usepackage[T1]{fontenc}
\usepackage{amsfonts}
\usepackage{nicefrac}
\usepackage{array}
\usepackage{scalerel}

\newcommand{\archmodel}[1]{\parbox[t]{0.19\textwidth}{\raggedright #1}}
\newcommand{\archparams}[1]{\parbox[t]{0.07\textwidth}{\raggedright #1}}
\newcommand{\archdesc}[1]{%
  \parbox[t]{\dimexpr0.74\textwidth-4\tabcolsep\relax}{\raggedright #1}}

\usepackage[colorlinks=true,
            linkcolor=blue!70!black,
            citecolor=green!50!black,
            urlcolor=blue!60!black,
            filecolor=blue!60!black]{hyperref}

\begin{document}

\title{Cross-Domain Transfer with Particle Physics Foundation Models:\\From Jets to Neutrino Interactions}

\author{Gregor Krzmanc}
\email{gregork@stanford.edu}
\affiliation{Department of Physics, Stanford University, Stanford, CA 94305, USA}

\author{Vinicius Mikuni}
\affiliation{Nagoya University, Kobayashi-Maskawa Institute, Aichi 464-8602, Japan}

\author{Benjamin Nachman}
\email{nachman@stanford.edu}
\affiliation{Department of Physics, Stanford University, Stanford, CA 94305, USA}
\affiliation{Fundamental Physics Directorate, SLAC National Accelerator Laboratory, Menlo Park, CA 94025, USA}

\author{Callum Wilkinson}
\affiliation{Lawrence Berkeley National Laboratory, Berkeley, CA 94720, USA}

\begin{abstract}
Future AI-based studies in particle physics will likely start from a foundation model to accelerate training and enhance sensitivity. As a step toward a general-purpose foundation model for particle physics, we investigate whether the OmniLearned and ParticleViT foundation models pretrained on diverse high-$Q^2$ simulated and real $pp$ and $ep$ collisions retain useful knowledge to a few-GeV fixed-target neutrino experiment.
We process MINERvA neutrino--nucleus scattering events and evaluate pretrained models on two types of tasks:
regression of available energy and binary classification of charged-current pion final states
($\mathrm{CC1\pi^{\pm}}$, $\mathrm{CCN\pi^{\pm}}$, and $\mathrm{CC1\pi^{0}}$).
Pretrained OmniLearned and ParticleViT models outperform similarly sized models trained from
scratch at the same compute budget, with the largest gains for OmniLearned on regression and for ParticleViT on classification.
When the same transformer architecture is instead initialized from unrelated text pretraining (BERT), this advantage appears only marginally for classification in terms of compute efficiency and not in any way for regression.
These results suggest that particle-level foundation models acquire inductive biases that generalize across large differences in energy scale, detector technology, and
underlying physics processes, pointing toward detector-agnostic inference in particle physics.
\end{abstract}

\maketitle

\section{Introduction}

Foundation models for language show promise for particle physics, especially for software development~\cite{Zhang:2024kws,gendreaudistler2025automatinghighenergyphysics}.  A growing number of models are being proposed to directly operate on particle physics data and not just natural language associated with these data~\cite{Feickert:2021ajf,Birk:2024knn,Mikuni:2024qsr,Hallin:2025ywf,Harris:2024sra,Golling:2024abg,Leigh:2024ked,Bardhan:2025icr,mikuni:2025tar,omnilearned,Park:2025ebs,Young:2025qah,Young:2025qbx, jan}.  Foundation models are large-scale models pretrained on broad datasets and subsequently
fine-tuned for diverse downstream tasks.  One foundation model that has been demonstrated to solve challenges across experiments is OmniLearned~\cite{Mikuni:2024qsr,mikuni:2025tar,omnilearned}, which is pretrained on particle-level jet representations from simulated and real $pp$ and $ep$ collider events. A model trained on similar data to OmniLearned, ParticleViT \cite{jan}, has been demonstrated to work better for many downstream jet tagging tasks.

In addition to its utility for collider physics tasks, OmniLearned has shown transferability to a variety of tasks, which are referred to as ``near'', ``medium'' and ``far'' transfers to indicate how different the application is from the initial training task. Near transfers have been studied over data fidelities (fast/full simulation), between experiments, and even across collision systems ($pp$ and $ep$). Far transfers have studied the utility of particle clouds --- point clouds of particles --- for non-particle data, applying the model to cosmological structure formation~\cite{omnicosmos} and even to molecular dynamics~\cite{omnimol}.  While it seems likely that there is a tradeoff between generality and effectiveness, these results suggest that particle-level transformers learn geometric and kinematic inductive biases that can be useful across a wide array of scientific contexts.

One aspect of transferability that is missing so far is a ``medium'' transfer. In this work we explore the model transfer from the high-energy collider regime (TeV-scale jets, $\mathcal{O}(10^2)$ particles per event, hermetic ($4\pi$) detectors) to a few-GeV fixed-target neutrino experiment (few-GeV neutrino interactions, $\mathcal{O}(1\text{--}10)$ reconstructed objects per event, asymmetric detector geometry). These regimes differ not just in detector technology but in the underlying physics and energy scale. Collider jets are dominated by quantum chromodynamics fragmentation \cite{Horgan_Jacob_1981}, while few-GeV neutrino interactions cover a complex transition region in which interactions can occur with an entire nucleus, a constituent nucleon within the nucleus and deep-inelastic scattering, with a variety of relatively low-energy nuclear effects further complicating the picture~\cite{Formaggio:2012cpf,Alvarez-Ruso:2014bla,Mosel:2016cwa,Katori:2016yel,NuSTEC:2017hzk,Mahn:2018mai,Jachowicz:2021ieb,SajjadAthar:2022pjt}. 

A major, multi-decade, experimental effort is underway focused on making precision neutrino oscillation measurements using few-GeV accelerator neutrino beams~\cite{Esteban:2024eli,ParticleDataGroup:2024cfk}. Accelerator neutrino beams are produced by directing intense proton beams onto fixed targets, focusing and sign-selecting secondary hadrons (primarily pions and kaons) of a desired energy, and allowing them to decay~\cite{Kopp:2006ky}. A key challenge for oscillation experiments is to disentangle the neutrino oscillation phenomenon, which varies as a function of neutrino energy and distance traveled, from uncertainties in both the rate at which neutrinos interact and in the energy distribution of neutrinos produced by the accelerator neutrino beam. In particular, because neutrino properties can only be reconstructed from the observable final-state interaction products, few-GeV neutrino--nucleus scattering uncertainties are affected by a large number of complex, relatively low-energy, nuclear effects~\cite{Formaggio:2012cpf,Alvarez-Ruso:2014bla,Mosel:2016cwa,Katori:2016yel,NuSTEC:2017hzk,Mahn:2018mai,Jachowicz:2021ieb,SajjadAthar:2022pjt}.


These neutrino experiments use detectors that simultaneously serve as the interaction medium, providing tracking and calorimetric measurement of the full hadronic final state down to the few-MeV level. Uncertainties in modeling these processes are large for current accelerator neutrino experiments~\cite{T2K:2011qtm, T2K:2025yoy, NOvA:2007rmc, NOvA:2025tmb, T2K:2025wet}, and are expected to be large, if not limiting, for next-generation experiments such as DUNE~\cite{Abi:2020wmh} and Hyper-Kamiokande~\cite{Hyper-Kamiokande:2018ofw}.

To tackle this challenge, experiments seek to directly measure neutrino--nucleus scattering cross sections by sampling accelerator neutrino beams close to the source, where oscillation effects can be neglected, both as part of oscillation experiments, and with dedicated experiments. The MINERvA experiment at Fermilab was specifically designed to address this challenge~\cite{minerva,MINERvA:2013zvz}, using a finely segmented scintillator tracking calorimeter and a variety of nuclear targets (C$_8$H$_8$, $^{12}$C, $^{56}$Fe, $^{208}$Pb, water and helium). MINERvA took data in the NuMI accelerator neutrino beam~\cite{numi}, which ran in both the low-energy (LE, 1--5 GeV)~\cite{MINERvA:2016iqn} and medium-energy (ME, 2--10 GeV)~\cite{MINERvA:2022vmb} beam configurations, with periods of muon neutrino and muon antineutrino enhanced beam running for each. Since 2012, the MINERvA collaboration has released a variety of neutrino--nucleus cross-section measurements which have significantly added to the understanding of few-GeV neutrino interactions by making detailed studies of the outgoing particle content and kinematic distributions~\cite{fiorentini,Fields_2017, MINERvA:2021csy,MINERvA:2021wjs,MINERvA:2022mnw,MINERvA:2022djk,MINERvA:2022esg,MINERvA:2022bno,MINERvA:2023kuz,MINERvA:2023ikp,MINERvA:2023ner,MINERvA:2025tem,MINERvA:2025hzq, MINERvA:2026apf}. Recently, MINERvA took a pioneering step in the accelerator neutrino community by making their data, simulated data, and analysis tools publicly available \cite{MINERvA:2025cfj}, which made the work described here possible.

The main contributions of this work are: (i)~defining ML-ready regression and classification tasks on publicly available MINERvA Open Data; (ii)~benchmarking several architectures (MLP, BDT and  transformer-based models) under a common event representation and training protocol; and (iii)~showing that pretraining on collider particle clouds transfers more usefully than text pretraining across this domain gap, supporting the idea that the models trained on HEP data acquire genuinely useful inductive biases for similar tasks.

This paper is organized as follows.  Section~\ref{sec:tasks} describes the regression and classification tasks that we use for studying transferability of OmniLearned and ParticleViT to neutrino simulated data from MINERvA.  The machine learning models, including our baselines, are introduced in Sec.~\ref{sec:methods}.  Numerical results are presented in Sec.~\ref{sec:results}.  The paper ends with conclusions and outlook in Sec.~\ref{sec:conclusions}.

\section{Tasks}
\label{sec:tasks}

We propose a regression task and three binary classification tasks on which we evaluate the models and their transfer-learning capabilities:
\begin{itemize}
    \item \textbf{Available hadronic energy regression.}
    In order to constrain different nuclear models at low energies, we aim to
    regress the \textit{available hadronic energy} for each event, $E_{\mathrm{available}}$.
    The available energy is defined as the sum of the kinetic energies of protons\footnote{Including, for completeness, antiprotons, although they are vanishingly rare at the relevant few-GeV energies.}
    plus the full energy of photons, pions, electrons, and charged kaons. All other mesons and baryons are neglected in our definition (although are rare).
    This definition explicitly excludes poorly modeled neutrons and nuclear binding effects,
    making it a quantity that can be reliably inferred from detector observables. 

    \item \textbf{Binary event classification.}
    We use three different positive class definitions, each constituting a separate task:
    $\mathrm{CC1\pi^{\pm}}$, $\mathrm{CCN\pi^{\pm}}$, and $\mathrm{CC1\pi^{0}}$.
    The positive classes consist of, respectively, charged-current (CC) events with
    exactly one charged pion ($\pi^{\pm}$), $N \geq 1$ charged pions, and exactly
    one neutral pion ($\pi^{0}$) with no charged pions.
    The negative class for each task consists of all events not in the positive class. The three classification tasks are derived from a single multi-class model.
\end{itemize}

These tasks are representative of the reconstruction challenges that few-GeV accelerator neutrino experiments attempt to solve, for both neutrino oscillation and neutrino cross-section analyses. Importantly, these are defined in terms of the {\it visible} quantities which an ideal detector would be able to reconstruct perfectly, avoiding explicit dependence on the input model used in the simulation --- a key challenge for few-GeV neutrino experiments. Available hadronic energy is the closest measurable quantity to true energy transfer to the nucleus, and is important for neutrino energy estimators used in neutrino oscillation experiments (by adding the reconstructed leptonic energy). Subdivision of events into different {\it observable} final-states helps to tease apart the complex nuclear physics problem that neutrino experiments need to untangle as initial interaction mechanisms and nuclear effects contribute differently to them.

\section{Methods}
\label{sec:methods}

\paragraph{Dataset.}
We use simulated MINERvA Medium Energy Forward Horn Current (FHC) events from
the standard Monte Carlo playlists 1A (both helium and water targets empty) and 1B (full helium target and empty water target).
Events are preprocessed into nested per-event tensors with truth labels and global features.
The data are split into training, validation, and test sets, stratified by interaction type.
The training dataset contains 6M events, while the validation and test sets each
contain 700k events.

\paragraph{Event representation.}
Each event is represented as a variable-length set of \emph{tokens}, one per
reconstructed object (a final-state particle captured by the detector, or a
cluster of energy deposits from one of the final-state particles).
The reconstructed objects include up to two photons, up to one muon, a variable
number of \emph{prongs} (electromagnetic energy deposits with a well-defined direction),
and a variable number of \emph{blobs} (calorimetric energy deposits without a
well-defined direction).
We follow the OmniLearned convention for continuous kinematic inputs: pseudorapidity
$\eta$ and azimuth $\phi$ (both defined with respect to the neutrino beam axis),
together with $\log p_{\mathrm{T}}/\text{GeV}$ and $\log E/\text{GeV}$.
A discrete particle-type label is stored as an integer channel alongside these four
kinematic scalars.
Beyond the OmniLearned kinematic block, each token carries 5 additional continuous features, and each event has an additional 15 global event-level continuous features.
See Appendix~\ref{app:dataset} for a complete description of the feature set.

\paragraph{Models.}
Particle tokens (and global event-level feature vectors) are processed with a transformer encoder.
We compare the following model families:

\begin{itemize}
    \item \textbf{MLP (global features):} A lightweight baseline ($\sim$218k parameters)
    with hidden width 128, 3 residual blocks, and sigmoid linear unit activation,
    operating only on the 15 global event-level features. This serves as a proxy for classical event reconstruction methods since only physics-engineered features are used.

    \item \textbf{BDT (global features):} a boosted decision tree model trained with the same inputs and against the same classification targets as the MLP. This serves as an additional proxy for classical event reconstruction.

    \item \textbf{ParticleViT:} A ViT-like transformer
    with additive positional encoding from $(\eta, \phi)$,
    evaluated at small (10M parameters) and medium (16M parameters) scales. The medium preset corresponds to ParticleViT-S~\cite{jan}.

    \item \textbf{OmniLearned:} The Particle Encoder Transformer v2 (PET2)~\cite{omnilearned}
    at small (3M parameters) and medium (52M parameters) presets.
    We use publicly available pretrained checkpoints. For the medium preset, we freeze the backbone and train only the task-specific heads due to compute constraints (OL-medium-frozen).

    \item \textbf{BERT:} A Bidirectional Encoder Representations from Transformers model trained on text data \cite{devlin2019bertpretrainingdeepbidirectional}. As inputs to the models are not permutation-invariant, we always use the same ordering: first the global event-level features, followed by muon, photons, energy-sorted prongs and energy-sorted blobs.

\end{itemize}

Randomly initialized counterparts of pretrained models (and the from-scratch MLP/BDT baselines) are marked with the suffix \texttt{-rw} (random weights).

Full architecture details are given in Appendix~\ref{app:arch}. Rough estimates of GPU time required for training of the models are given in Appendix \ref{app:compute}. 

\paragraph{Training.}
Models are trained using the Adam optimizer~\cite{kingma2017adammethodstochasticoptimization} with a learning rate of $10^{-4}$ and a batch size of
2048 events.
Separate models are trained for classification and regression.
The regression model minimizes the Smooth L1 Loss \cite{smoothl1} applied to $\log(1 + E_\mathrm{available})$,
and the classification model minimizes the cross-entropy loss with class weights
proportional to inverse class frequency to handle class imbalance. The classifier model is trained to predict one of 5 classes: $CC1\pi^\pm$, $CC1\pi^0$, $CCN\pi^{\pm}$ ($N > 1$), any other charged-current event, and any neutral-current event. The signal and background classes are then selected separately for each use case.

Models are evaluated on the validation set every 1000 steps; the checkpoint with
the best validation loss is selected as the final model.
All results are reported over 4 independent runs.

\section{Results}
\label{sec:results}

\subsection{Computational efficiency}

\begin{figure*}[t]
    \centering
    \includegraphics[width=\textwidth]{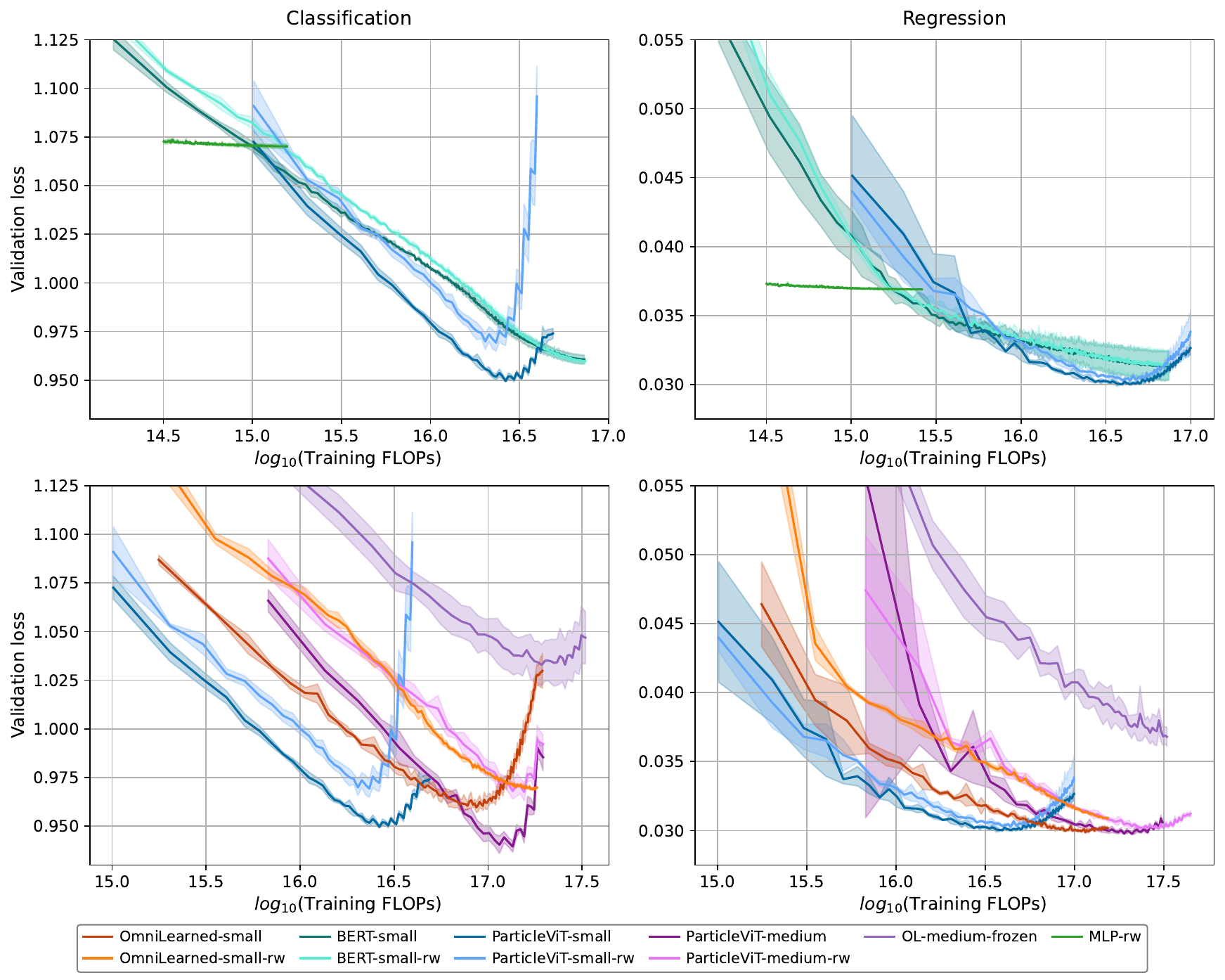}
    \caption{Validation loss vs.\ number of cumulative training FLOPs for classification
    tasks (left) and the regression task (right), using the MINERvA Open Data playlist 1A.
    The top row compares BERT-small, ParticleViT-small and the MLP baseline, while the bottom row compares all the models pretrained on jets.
    Pretrained models on jet data (OmniLearned-small and ParticleViT models) achieve lower loss at equivalent compute budgets compared to the same scratch-trained models (OmniLearned-small-rw, ParticleViT-small-rw and ParticleViT-medium-rw models). Pre-trained ParticleViT-small model achieves its lowest validation loss in around a third of the compute required for OmniLearned-small both for classification and regression.
    Validation loss is reported every 1000 steps.
    For some models, the curves are shown only from a later point on, as the validation loss has a high variance early in training.}
    \label{fig:flops}
\end{figure*}

Figure~\ref{fig:flops} shows validation loss as a function of training floating-point operations (FLOPs)
for both classification and regression.
Comparing OmniLearned-small, ParticleViT-small, and ParticleViT-medium, with their randomly initialized counterparts ($\texttt{-rw}$ suffix) isolates the effect of pretraining from architecture differences: the pretrained model reaches a lower validation loss at a given FLOP budget.
Comparing OmniLearned and ParticleViT models, the latter ones also outperform the former both in terms of compute efficiency and final classification performance.
OmniLearned models perform slightly better than ParticleViT models in the regression task in the low-$q_3$ region.

ParticleViT is pretrained solely for jet tagging (classification), whereas OmniLearned is pretrained jointly for jet tagging and generation~\cite{omnilearned,jan}. ParticleViT matches or exceeds OmniLearned across MINERvA classification tasks, which suggests that supervised jet tagging alone is already a particularly effective pretraining prior.

Performing the same study using a text-trained language model shows that the same patterns only marginally hold for the classification task and do not hold for the regression task.
A similar study with pretrained LLM weights on cosmological maps found a clearer benefit from language-model initialization~\cite{Heneka_2026}, which is not observed here.

See also Appendix \ref{app:perf} for the corresponding comparison as a function of training steps, showing that pretrained models (OmniLearned-small) reach a lower validation loss at the same number of training steps compared to non-pretrained models.

\subsection{Classification performance}

\begin{figure*}[t]
    \centering
    \includegraphics[width=\linewidth]{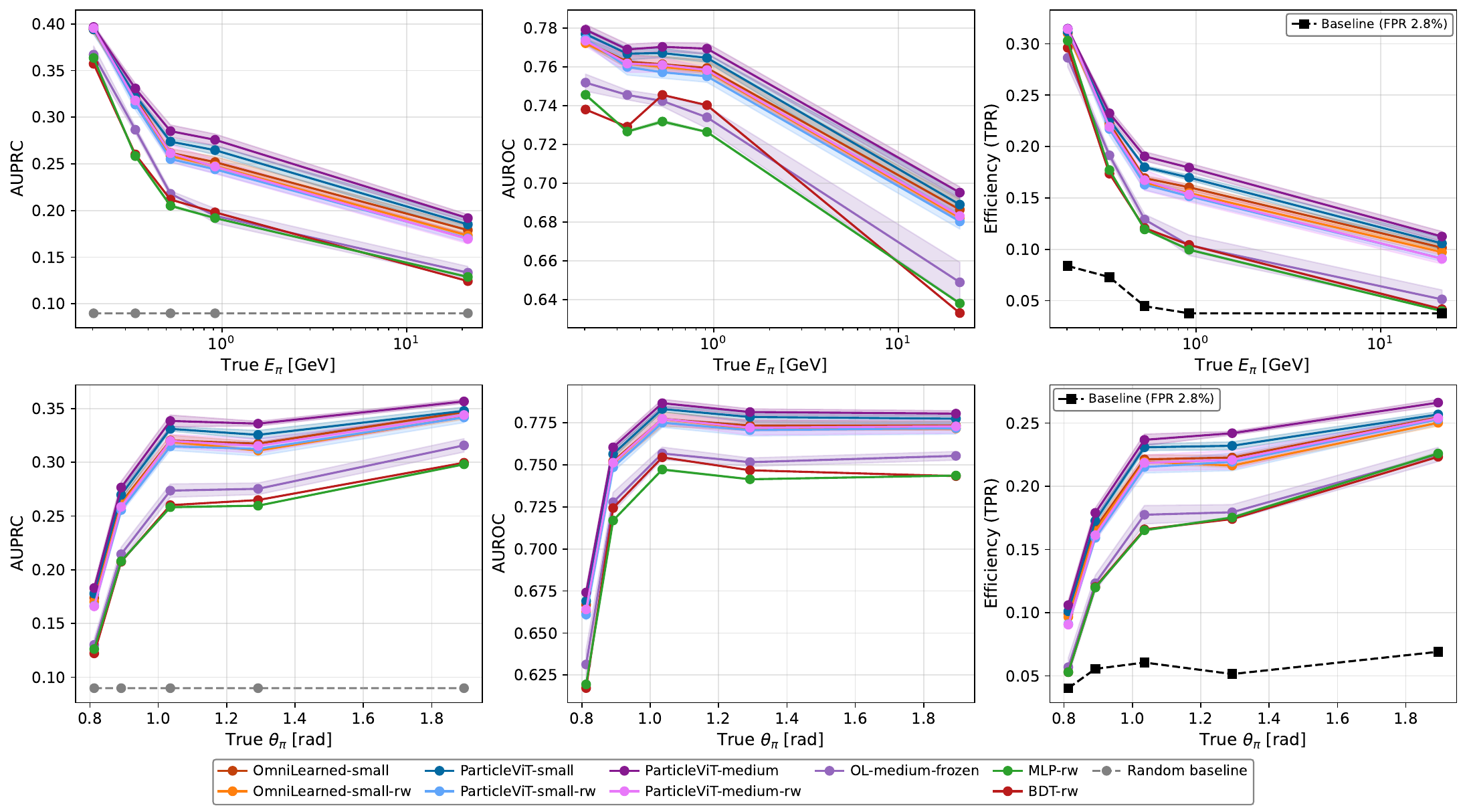}
    \caption{Discrimination metrics for the $\mathrm{CC1\pi^{\pm}}$ tagging task as a function of true pion kinematics: area under the precision--recall curve (AUPRC, left column), area under the ROC curve (AUROC, center), and true positive rate at a fixed false positive rate (right column). The top row uses bins in true pion energy; the bottom row uses bins in true pion polar angle. In each bin, positives are true signal events falling in that bin, while the negative sample is the \emph{full} background, held fixed across bins so that only the in-bin signal population changes. TPR is computed at the baseline false-positive rate matched globally on the full test set; see Fig.~\ref{fig:tpr_perbin} for the per-bin matched comparison.}
    \label{fig:cc1pipm}
\end{figure*}

\begin{figure*}[t]
    \centering
    \includegraphics[width=\linewidth]{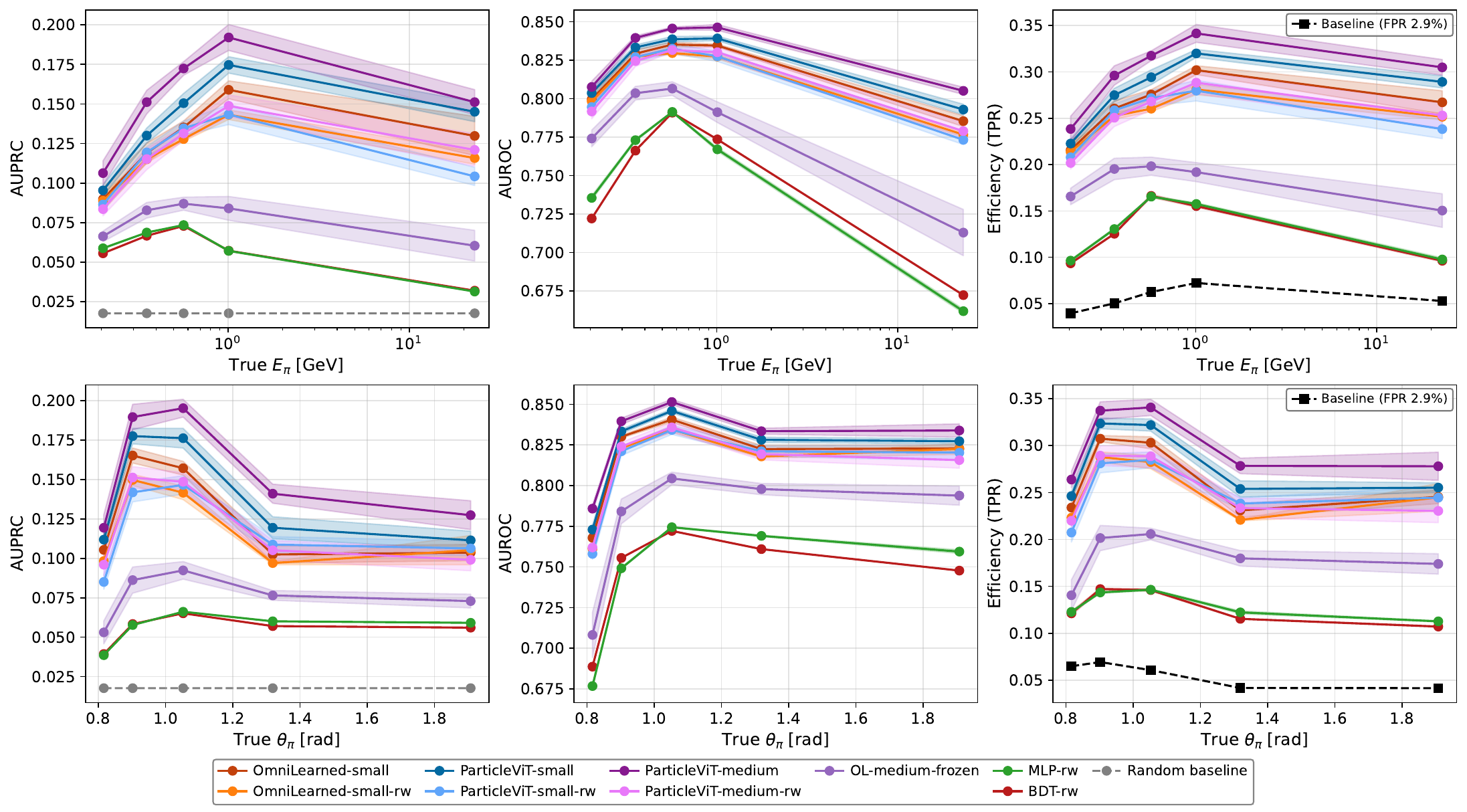}
    \caption{Discrimination metrics for the $\mathrm{CC1\pi^{0}}$ tagging task as a function of true pion kinematics: area under the precision--recall curve (AUPRC, left column), area under the ROC curve (AUROC, center), and true positive rate at a fixed false positive rate (right column). The top row uses bins in true pion energy; the bottom row uses bins in true pion polar angle. In each bin, positives are true signal events falling in that bin, while the negative sample is the \emph{full} background, held fixed across bins so that only the in-bin signal population changes. TPR is computed at the baseline false-positive rate matched globally on the full test set; see Fig.~\ref{fig:tpr_perbin} for the per-bin matched comparison.}
    \label{fig:cc1pi0}
\end{figure*}

\begin{figure*}[t]
    \centering
    \includegraphics[width=\linewidth]{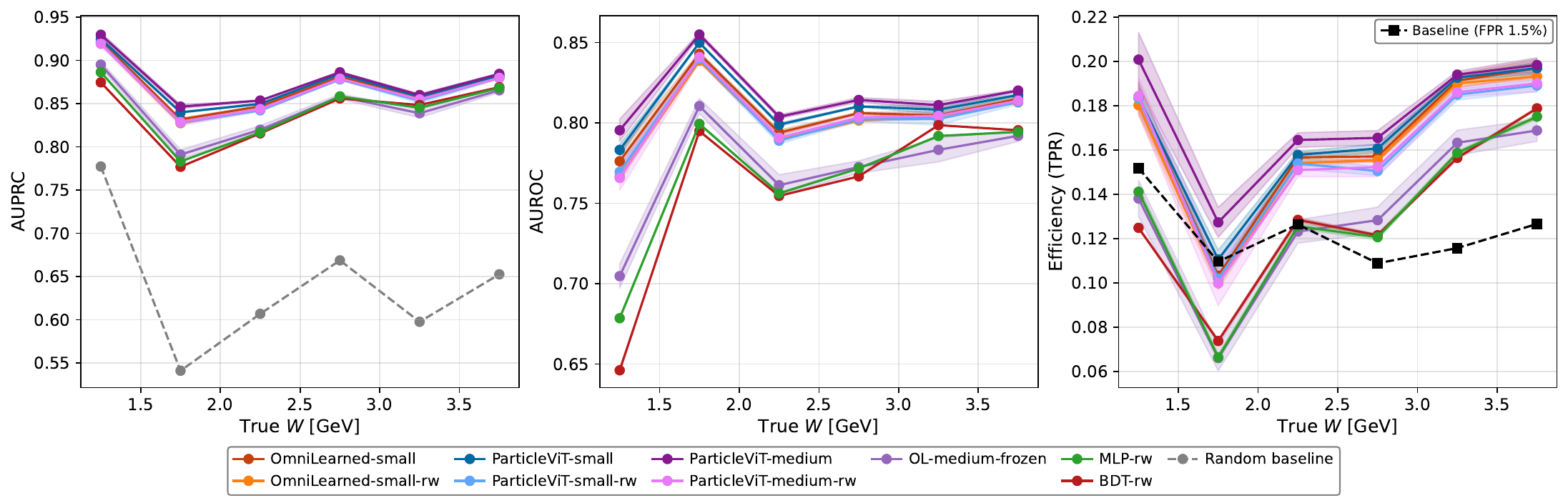}
    \caption{Discrimination metrics for the $\mathrm{CCN\pi^{\pm}}$ tagging task as a function of invariant mass of the hadronic system $W$: area under the precision--recall curve (AUPRC, left column), area under the ROC curve (AUROC, center), and true positive rate at a fixed false positive rate (right column). Efficiency is computed using the baseline's false-positive rate matched globally on the full test set; see Fig.~\ref{fig:tpr_perbin} for the per-bin matched comparison.
    }
    \label{fig:ccnpi}
\end{figure*}

Figures~\ref{fig:cc1pipm}--\ref{fig:ccnpi} show the classification performance
for each of the three tasks as true positive rate (TPR) evaluated at the false
positive rate (FPR) of a classical cut-based baseline (see Appendix \ref{app:baselines}), binned in physically
motivated kinematic variables.
Unless noted otherwise, the operating point is chosen once on the full test set so that the model FPR matches the baseline FPR globally; TPR is then reported in each kinematic bin at that fixed threshold.
For the single-pion tasks ($\mathrm{CC1\pi^{\pm}}$ and $\mathrm{CC1\pi^{0}}$),
signal events are binned in pion energy and angle; for the multi-pion task ($\mathrm{CCN\pi^{\pm}}$),
events are binned in invariant mass of the hadronic system $W$.
Signal and background composition in these variables is shown in Appendix~\ref{app:dataset}.

Pretrained ParticleViT and OmniLearned models show marginally improved metrics across most bins in all three tasks compared to their randomly-initialized counterparts and other models.
For $\mathrm{CC1\pi^{\pm}}$ (Fig.~\ref{fig:cc1pipm}), the gain is consistent across most of the pion energy range.
A qualitatively similar pattern holds for $\mathrm{CC1\pi^{0}}$ (Fig.~\ref{fig:cc1pi0}),
where neutral-pion identification adds the additional challenge of reconstructing
photon pairs.
For the multi-pion task $\mathrm{CCN\pi^{\pm}}$ (Fig.~\ref{fig:ccnpi}), the pretrained ParticleViT models show marginal gain at low $W$.
With the global FPR-matched threshold, some models fall below the cut-based baseline in individual bins: a global cut can allocate background differently across kinematic regions than the cut-based selection, so bin-wise TPR comparisons at fixed global FPR are not always favorable even when ranking metrics (AUPRC, AUROC) remain strong.

\begin{figure*}[t]
    \centering
    \includegraphics[width=\linewidth]{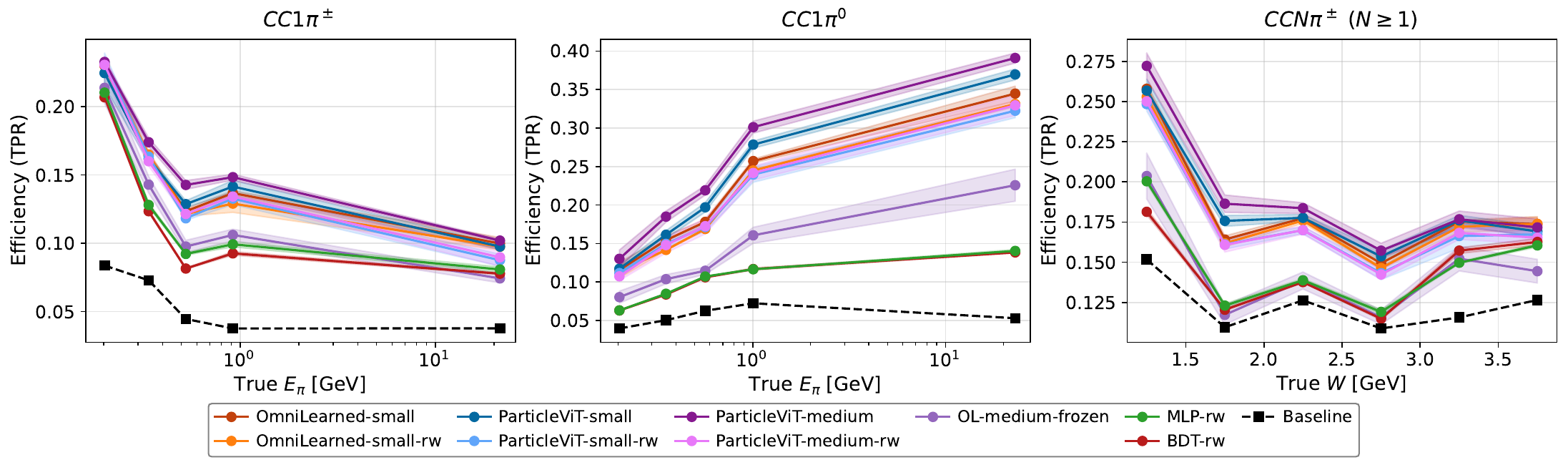}
    \caption{Event classification efficiency at the cut-based baseline's false-positive rate, matched independently in each kinematic bin, for $\mathrm{CC1\pi^{\pm}}$ and $\mathrm{CC1\pi^{0}}$ (vs.\ true pion energy $E_\pi$) and $\mathrm{CCN\pi^{\pm}}$ with $N\geq 1$ (vs.\ true hadronic invariant mass $W$).
    Shaded bands show the spread over four independent training runs.
    Unlike the globally matched TPR panels in Figs.~\ref{fig:cc1pipm}--\ref{fig:ccnpi}, every neural model remains above the cut-based baseline in all bins.}
    \label{fig:tpr_perbin}
\end{figure*}

Figure~\ref{fig:tpr_perbin} therefore repeats the efficiency comparison with the baseline FPR matched \emph{independently in each bin}.
Under this per-bin operating point, all models outperform the cut-based baseline across the full kinematic range for all three tasks, confirming that the occasional underperformance in Figs.~\ref{fig:cc1pipm}--\ref{fig:ccnpi} is an artifact of global thresholding rather than a loss of discrimination power.

See Appendix \ref{app:perf_bert} for a comparison of classification performance of the fully-trained BERT models: the final performance for pretrained and randomly-initialized BERT models is the same, contrary to ParticleViT and OmniLearned.
 
\subsection{Regression performance}

\begin{figure}
    \centering
    \includegraphics[width=\columnwidth]{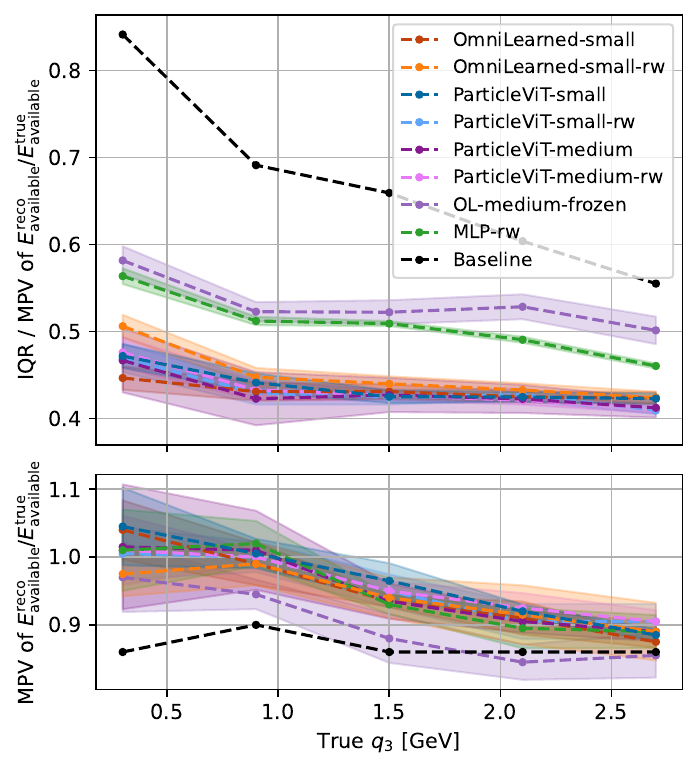}
    \caption{Interquartile range (top) and most probable value (bottom) of the
    available hadronic energy residual distribution vs.\ $q_3$, for playlist 1A.
    Smaller IQR indicates better energy resolution; MPV near unity indicates
    no systematic bias. Events with $E^\mathrm{true}_\mathrm{available} < 0.05$~GeV are excluded.}
    \label{fig:iqr_1A}
\end{figure}

Figure~\ref{fig:iqr_1A} shows the available hadronic energy regression performance on
playlist 1A as the interquartile range (IQR) of the residual ratio distribution as a function of the three-momentum transfer $q_3$, together with the most probable value (MPV) of the
residual ratio.
Events with $E^\mathrm{true}_\mathrm{available} < 0.05$~GeV are excluded from all regression comparisons: the ratio $E^\mathrm{reco}_\mathrm{available}/E^\mathrm{true}_\mathrm{available}$ is ill-behaved as the denominator approaches zero, so a handful of near-zero-energy events would otherwise dominate the IQR and MPV without reflecting any difference in reconstruction quality.
Pretrained OmniLearned models (OmniLearned-small) achieve a narrower IQR in the lowest $q_3$ bin,
indicating more precise energy reconstruction; however, there is some systematic bias (MPV slightly below unity) across all models at higher $q_3$. The relative ordering of models is preserved for OmniLearned models; however, pre-training seems to have little-to-no impact on the final performance of the ParticleViT models.

\begin{figure}
    \centering
    \includegraphics[width=\columnwidth]{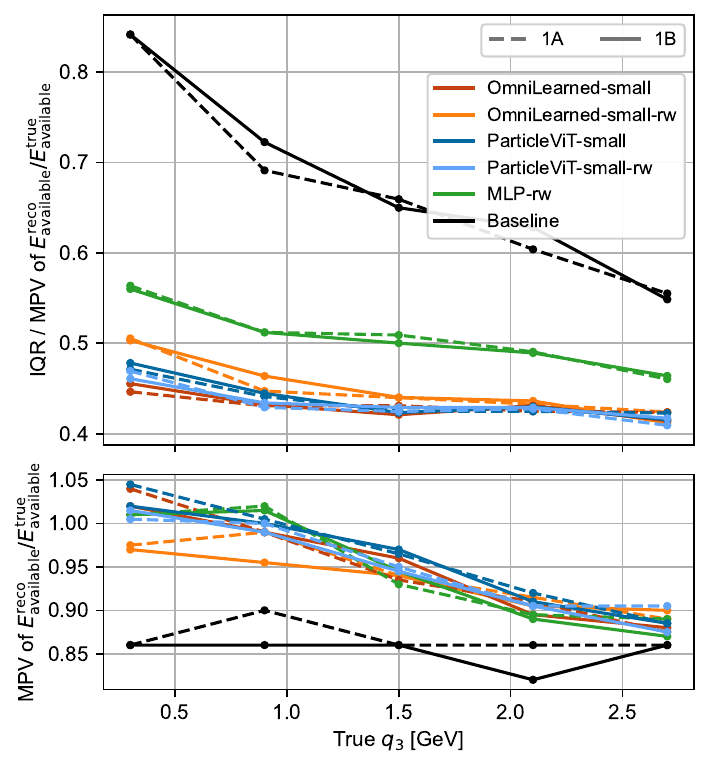}
    \caption{Available hadronic energy resolution summary for playlists 1A and 1B.
    The relative ordering of models is consistent with the single-playlist result
    in Fig.~\ref{fig:iqr_1A}.}
    \label{fig:iqr_1A1B}
\end{figure}

Figure~\ref{fig:iqr_1A1B} extends the comparison to both playlists 1A and 1B.

\begin{figure*}[t]
    \centering
    \includegraphics[width=\textwidth]{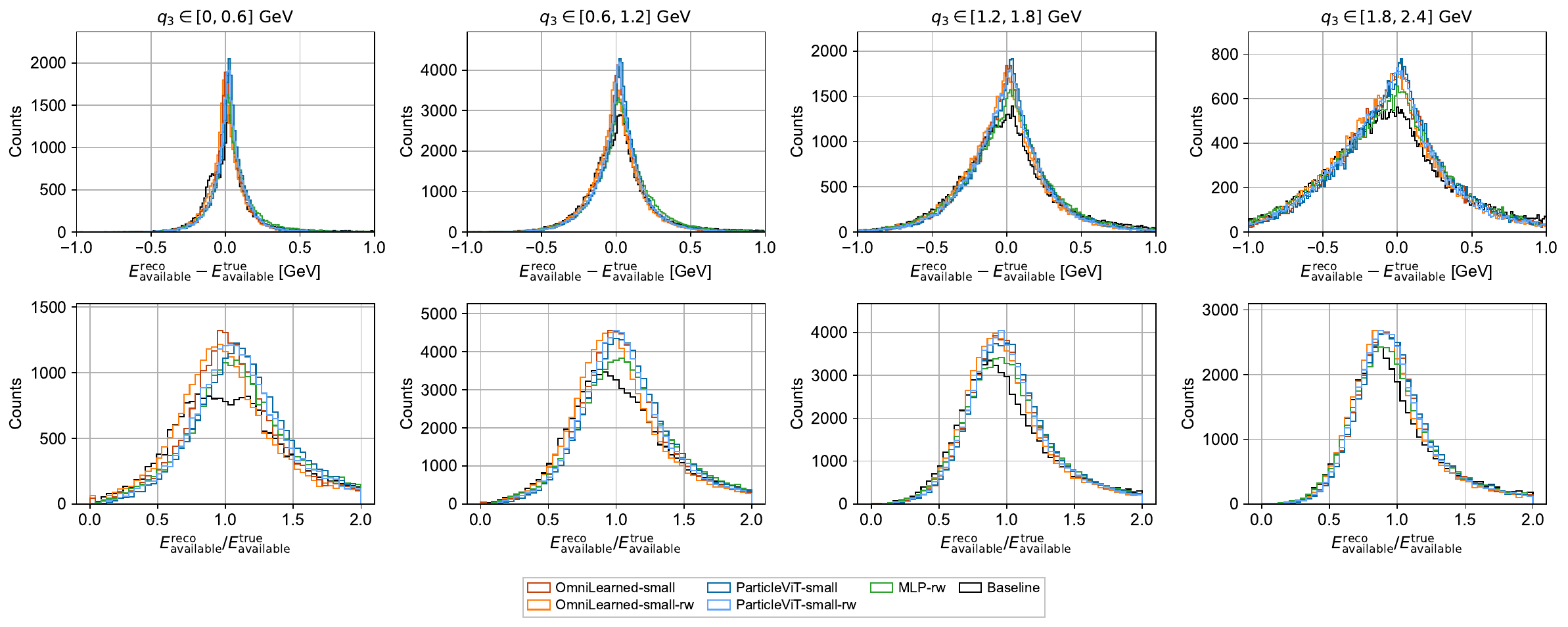}
    \caption{Available hadronic energy residual distributions in four bins of $q_3$
    for muon-selected events (top row: signed residual $E^\mathrm{reco}_\mathrm{available}-E^\mathrm{true}_\mathrm{available}$;
    bottom row: ratio $E^\mathrm{reco}_\mathrm{available}/E^\mathrm{true}_\mathrm{available}$).
    Pretrained OmniLearned models (OmniLearned-small) produce narrower, more symmetric distributions compared to scratch-trained models and ParticleViT models,
    especially at low $q_3$. Events with $E^\mathrm{true}_\mathrm{available} < 0.05$~GeV are excluded.}
    \label{fig:residuals}
\end{figure*}

Figure~\ref{fig:residuals} shows the full residual distributions in four $q_3$
bins for muon-selected events, allowing a more detailed view of the shape of the
reconstruction error.
All machine learning models produce narrower, more symmetric residual distributions compared to the baseline,
particularly in the lowest $q_3$ bin, where nuclear effects are largest and
the reconstruction is most challenging.

\section{Conclusions}
\label{sec:conclusions}

We have shown that inductive biases learned during pretraining on collider
physics jets transfer to the qualitatively different setting of a few-GeV fixed-target neutrino experiment.
Pretrained OmniLearned and ParticleViT models outperform the same-architecture models trained from
scratch in compute efficiency, with larger final-performance gains for OmniLearned on available-energy regression and for ParticleViT on pion-final-state classification.
Because ParticleViT was pretrained with jet tagging alone, while OmniLearned combined tagging with generation, the strong ParticleViT transfer also suggests that classification on jets is a particularly good pretraining objective.
By contrast, initializing from a text-pretrained BERT checkpoint yields at most a marginal compute-efficiency benefit for classification and no benefit for regression, indicating that the useful
transferred structure is domain-specific rather than a generic transformer prior.

This contrast is unsurprising from a machine-learning perspective: priors learned on a related modality (here, sparse particle kinematics) are more likely to help than unrelated ones, and the benefit of pretraining degrades as the source and target domains diverge~\cite{Yosinski:2014,Raghu:2019}.
What is more surprising is the scale of the domain gap across which the particle-physics prior still helps.
Collider jet physics and MINERvA neutrino scattering differ in energy scale
(TeV vs.\ few GeV), detector technology (calorimetric towers and a tracker with nearly full $4\pi$
coverage vs.\ a segmented scintillator calorimeter with beam-axis asymmetry),
multiplicity ($\mathcal{O}(10^2)$ particles per jet event vs.\ $\mathcal{O}(1\text{--}10)$
reconstructed objects per MINERvA event), and the underlying physical processes
(QCD fragmentation vs.\ a variety of interactions plus a number of low-energy
nuclear effects).
Yet the learned particle-level representations retain sufficient generality to
confer a performance advantage.
We interpret this as evidence that pretraining models on jets instills a useful geometric and
kinematic prior that is broadly applicable across particle
physics domains.

Our results suggest that pretraining on jets may reduce the amount of compute
required in new experimental settings, which could be particularly valuable
during detector commissioning or rapid hypothesis testing.
More broadly, we view this as a step toward \emph{detector-agnostic inference}:
a regime where a single foundation model can be adapted to new experiments
with minimal retraining.
Concretely, pretrained models could enable (i)~rapid optimization of new
experimental designs by reducing the amount of compute required to evaluate
performance, and (ii)~rapid testing of new physics hypotheses in established
experiments by lowering the cost of training specialized classifiers.

Several directions remain for future work. A new model could be created that includes neutrino events directly in its pretraining task. This would likely improve efficacy for MINERvA and it would be interesting to explore its potential utility for other neutrino detectors such as MicroBooNE and DUNE.


\section*{Code Availability}

The code for this paper can be found at \url{https://github.com/gregorkrz/minerva-ml}.  The data are available at \url{https://minerva.fnal.gov/opendata/} \cite{MINERvA:2025cfj}.

\begin{acknowledgments}
The authors thank the MINERvA Collaboration for making their data, simulated data, and analysis tools available to the community and for their support in working with the dataset. Specifically, we thank MINERvA collaborators Oscar Moreno-Palacios, David Last, and Rik Gran for their technical assistance and discussion. We also thank Sam Young and Jan-Lucas Uslu for helpful discussions. BN is supported by the U.S. Department of Energy (DOE), Office of Science under contract DE-AC02-76SF00515. VM is supported by JST EXPERT-J, Japan Grant Number JPMJEX2509. CW is supported by the U.S. Department of Energy (DOE), Office of Science under contract DE-AC02-05CH11231. This research used resources of the National Energy Research Scientific Computing Center, a DOE Office of Science User Facility supported by the Office of Science of the U.S. Department of Energy under Contract No. DE-AC02-05CH11231 using NERSC awards HEP-ERCAP0035546. 
\end{acknowledgments}
\clearpage
\bibliography{references}

\appendix

\section{Dataset}\label{app:dataset}

Each event is represented using 15 global features and a set of particle tokens (reconstructed objects).

Each particle token carries a 10-dimensional feature vector: pseudorapidity $\eta$, azimuth $\phi$, $\log(p_{\mathrm{T}}+\epsilon)$, $\log(E+\epsilon)$, an integer PID (node type) label,  $\log ( \mathrm{mean\ } \mathrm{d}E/\mathrm{d}x + \epsilon)$ (set to zero for muons and calorimetric blobs), spatial coordinates $(x,y,z)$ and time $t$, with zeros used where a channel is not defined for that object type. The first three kinematic features are computed from the neutrino beam axis to match the format of inputs to OmniLearned models. The list includes at most one MINOS-matched muon, up to two reconstructed photons from $\pi^0\to\gamma\gamma$, up to $20$ calorimetric blobs and $10$ track-based prongs (pion, electromagnetic-shower, or muon-like hypotheses), with lower-energy excess blobs and prongs merged into single aggregated nodes. There are at most 33 objects per event.

The global features summarize the event using aggregated and event-level quantities: calorimetric energy-related quantities (\verb|muon_fuzz_energy|, \verb|muon_iso_blobs_energy|) and passive recoil energy (\texttt{part\_\allowbreak response\_\allowbreak total\_\allowbreak recoil\_\allowbreak passive\_\allowbreak allNonMuonClusters\_\allowbreak id},
\texttt{part\_\allowbreak response\_\allowbreak total\_\allowbreak recoil\_\allowbreak passive\_\allowbreak allNonMuonClusters\_\allowbreak od},
and the sum of both), and quantities that are used in cut-based classification baselines (Michel electron count, number of reconstructed muons, diphoton invariant mass $m_{\gamma\gamma}$ when there are two reconstructed photons, charged-pion prong count). The remaining entries are log-scaled totals of reconstructed energy summed over different node types.

A complete description of the feature set used for MINERvA event representation
is provided at \href{https://github.com/gregorkrz/minerva-ml/blob/master/DATASET.md}{https://github.com/gregorkrz/minerva-ml/blob/master/DATASET.md}.

The models' performance is evaluated in bins of truth quantities derived from the final-state particle branches. The invariant mass of the hadronic system $W$ is computed from all the final-state particles except leptons. The three-momentum transfer is defined as
\begin{equation}\label{eqn:q3}
    q_3=\sqrt{\bigl(2E_\nu(E_\mu-|\vec{p}_\mu|\cos\theta_\mu)-m_\mu^2\bigr)+(E_\nu-E_\mu)^2}.
\end{equation}
Three-momentum transfer from Eq.~\eqref{eqn:q3} is defined only for events with a muon in the final state.

The single-pion final states ($CC1\pi^\pm$, $CC1\pi^0$) are binned in terms of the true angle and energy of the pion ($\pi^\pm$ or $\pi^0$).

\begin{figure*}[t]
    \centering
    \includegraphics[width=\textwidth]{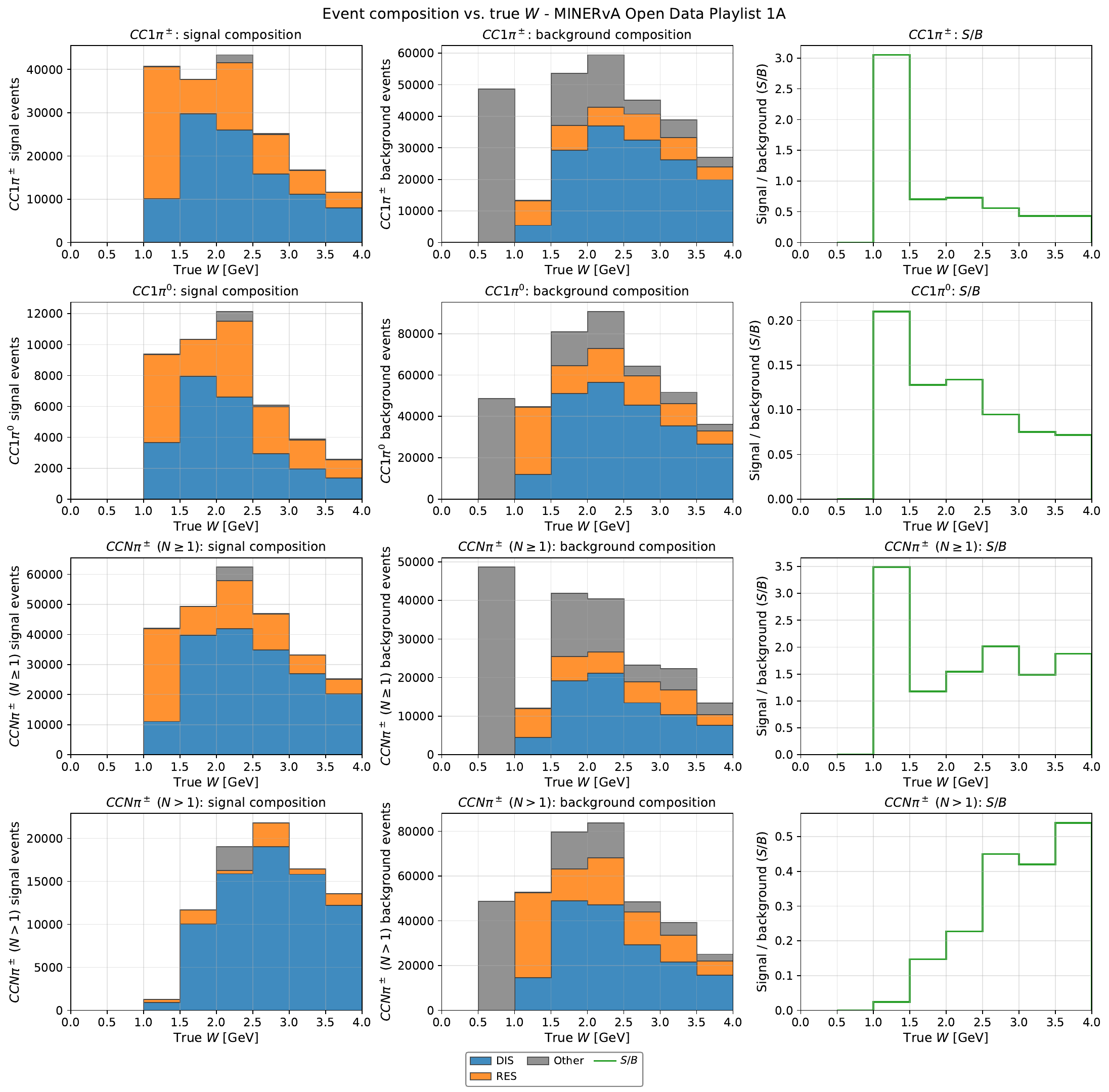}
    \caption{Event composition in bins of true hadronic invariant mass $W$ for playlist~1A.
    Each row corresponds to a different signal definition.
    Left: signal counts by interaction category.
    Center: background counts by interaction category.
    Right: signal-to-background ratio.}
    \label{fig:eventcomp_W}
\end{figure*}

\begin{figure*}[t]
    \centering
    \includegraphics[width=\textwidth]{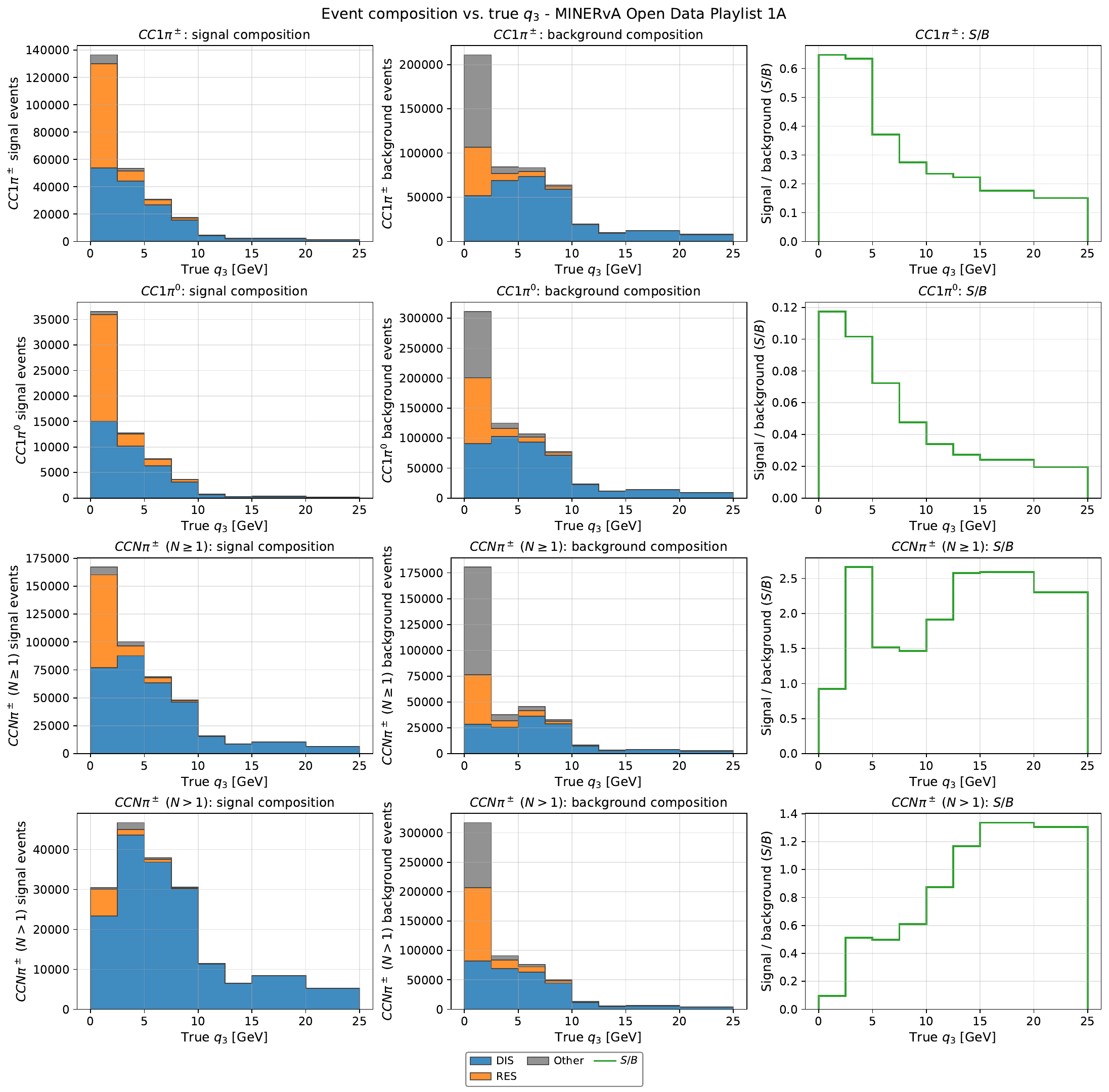}
    \caption{Same as Fig.~\ref{fig:eventcomp_W}, but reported in bins of the true three-momentum transfer $q_3$.}
    \label{fig:eventcomp_q3}
\end{figure*}

\begin{figure*}[t]
    \centering
    \includegraphics[width=\textwidth]{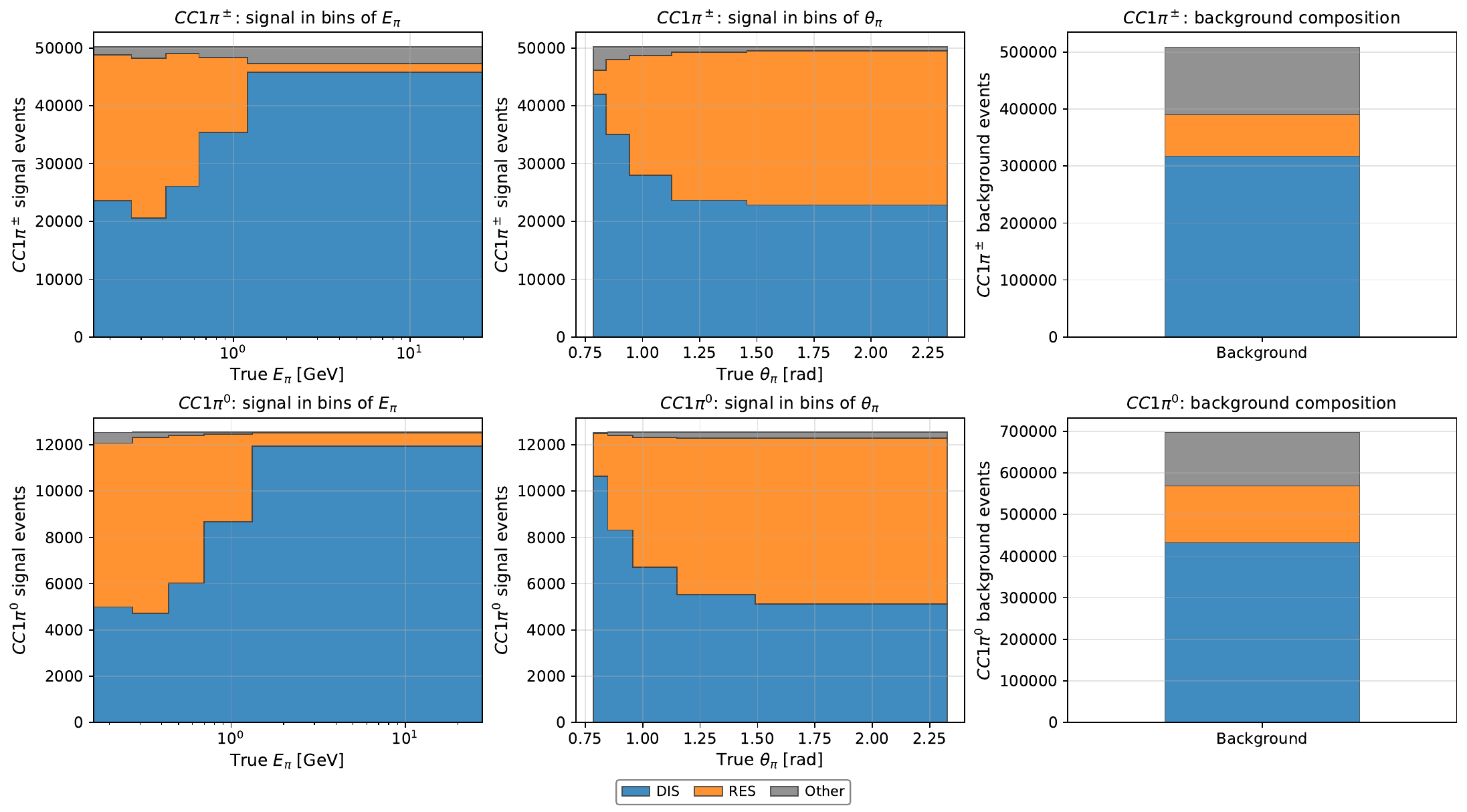}
    \caption{Composition of the single-pion signals versus true pion energy $E_\pi$ and polar angle $\theta_\pi$, together with the corresponding background composition, for playlist~1A.}
    \label{fig:eventcomp_pion}
\end{figure*}

Figures~\ref{fig:eventcomp_W}--\ref{fig:eventcomp_pion} show the signal and background composition of playlist~1A in these variables, stratified by GENIE interaction category (deep inelastic scattering, resonance production, and other). For $\mathrm{CCN\pi^{\pm}}$, both the inclusive ($N\geq 1$) and exclusive multipion ($N>1$) selections used in training are shown.

\section{Model architecture}\label{app:arch}
Table~\ref{tab:architectures} summarizes the architecture of each model evaluated.

\begin{table*}[t]
  \centering
  \small
  \caption{Model configurations used in the experiments. Randomly-initialized models are marked with \texttt{-rw}.}
  \label{tab:architectures}
  \renewcommand{\arraystretch}{1.15}
  \begin{tabular}{@{}lll@{}}
    \toprule
    Model & \# params & Architecture summary \\
    \midrule
    \archmodel{OmniLearned-small} &
    \archparams{3M} &
    \archdesc{Particle-encoder transformer (PET2), as described in
      \cite{omnilearned}.} \\
    \addlinespace[0.3em]
    \cmidrule(l){1-3}

    \archmodel{OmniLearned-medium (OL-medium-frozen)} &
    \archparams{52M} &
    \archdesc{Particle-encoder transformer (PET2) \cite{omnilearned} with
      frozen backbone ($\sim$21M trainable parameters).} \\
    \addlinespace[0.3em]
    \cmidrule(l){1-3}

    \archmodel{ParticleViT-small} &
    \archparams{9.7M} &
    \archdesc{ParticleViT (embedding variant): width 448, 4 transformer
      layers, 7 attention heads; SwiGLU MLP ratio $8/3$, RMSNorm, QK-Norm;
      CLS readout. Pretrained vs.\ random init (\texttt{-rw}).} \\
    \addlinespace[0.3em]
    \cmidrule(l){1-3}

    \archmodel{ParticleViT-medium} &
    \archparams{16M} &
    \archdesc{ParticleViT: width 512, 5 transformer layers, 8 attention
      heads; SwiGLU MLP ratio $8/3$, RMSNorm, QK-Norm; CLS readout.
      Pretrained vs.\ random init (\texttt{-rw}). Corresponds to the
      ParticleViT model from \cite{jan}.} \\
    \addlinespace[0.3em]
    \cmidrule(l){1-3}

    \archmodel{BERT-small} &
    \archparams{4.4M} &
    \archdesc{Hugging Face \texttt{prajjwal1/bert-tiny} encoder without
      tokenization \cite{berttiny_cite}: per-particle features and global
      features are linearly projected in the model's embedding space and
      outputs undergo mean-pooling.} \\
    \addlinespace[0.3em]
    \cmidrule(l){1-3}

    \archmodel{MLP (global features)} &
    \archparams{218k} &
    \archdesc{Using global features only; hidden width 128, 3 residual
      blocks, sigmoid linear unit activation.} \\
    \addlinespace[0.3em]
    \cmidrule(l){1-3}

    \archmodel{BDT (global features)} &
    \archparams{---} &
    \archdesc{Gradient boosting on global features; 400 boosting iterations,
      max 63 leaf nodes, learning rate 0.1.} \\

    \bottomrule
  \end{tabular}
\end{table*}

\section{Compute requirements}\label{app:compute}

The models are trained on a single NVIDIA A100 GPU. Training of the classification models takes approximately 6 hours for OmniLearned-small, approx. 1.5 hours for ParticleViT-small, and approximately 2 hours for ParticleViT-medium. Training of the regression models takes approximately 8 hours for OmniLearned-small, approx. 1.5 hours for ParticleViT-small, and approx. 2.5 hours for ParticleViT-medium. Training of all BERT models takes approx. 2 hours.

\section{Performance of the fully-trained text models}\label{app:perf_bert}

\begin{figure*}[t]
    \centering
    \includegraphics[width=\textwidth]{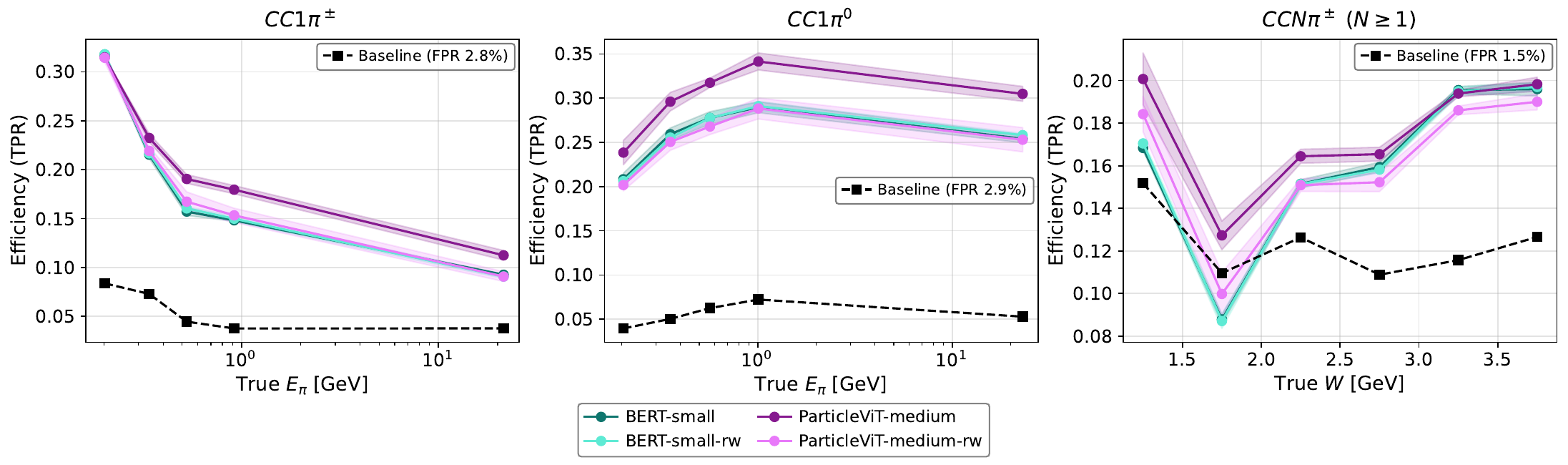}
    \caption{Efficiency of the event classifier for discriminating signal from background at matched false-positive rate of the cut-based baseline, for ParticleViT-medium and BERT-small models as well as their randomly-initialized counterparts (\texttt{-rw}). Starting training from a pre-trained BERT model seems to offer no advantage to final performance of the model.}
    \label{fig:bert_physics_performance}
\end{figure*}

Figure \ref{fig:bert_physics_performance} shows that  performance of the fully-trained classification model stays the same, no matter whether pretrained BERT weights are used or not.

\section{Performance vs.\ training step}\label{app:perf}

\begin{figure*}[t]
    \centering
    \includegraphics[width=\textwidth]{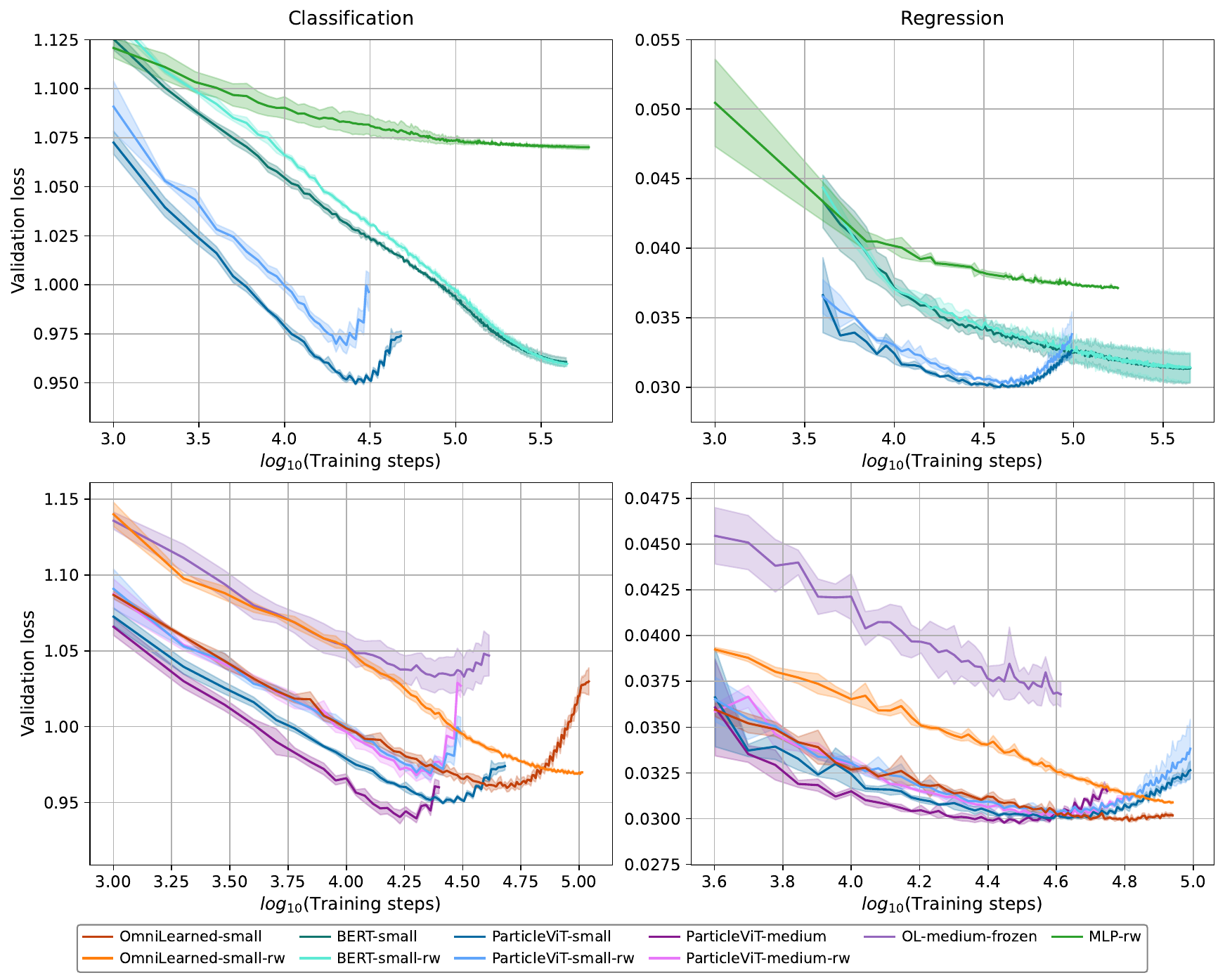}
    \caption{Validation loss vs.\ training step for classification tasks (left) and regression tasks (right).
    The top row compares BERT-small, ParticleViT-small and the MLP baseline, while the bottom row compares the models pretrained on jets.
    The pretrained model ParticleViT-small reaches its lowest validation loss in approx. 40-50\% fewer steps than pretrained OmniLearned-small model both for classification and regression.
    For some models, the curves are shown only from a later point on, as the validation loss has a high variance early in training.}
    \label{fig:log_steps_classification_regression}
\end{figure*}

Figure \ref{fig:log_steps_classification_regression}
shows validation loss as a function of training steps for classification and
regression tasks, respectively.
Pretrained OmniLearned models converge to lower validation loss more quickly
than models trained from scratch, consistent with the FLOP-efficiency results
in Fig.~\ref{fig:flops}.

\section{Evaluation baselines}
\label{app:baselines}

\subsection{Regression: $E_{\mathrm{available}}$}
\label{app:baselines:regression}

The black curves labeled \emph{baseline} in the $q_{3}$-binned regression figures are not a learned model: they use an estimate of available energy from the standard MINERvA analysis chain. The quantity used as the baseline is \texttt{blob\_recoil\_E}, multiplied by a scale factor of $1.17$, following the MINERvA low-recoil available-energy definition~\cite{MATMINERvA_LowRecoilFunctions}.

\subsection{Classification: interaction tagging}
\label{app:baselines:classification}

The classification figures show two distinct reference quantities.

\paragraph{Random (signal-fraction) baseline.} The baseline is not physically meaningful; it represents the AUPRC score of a random model for each bin (which coincides with fraction of signal events). 

\paragraph{Reconstruction baseline.}
The black dashed curves (square markers) on the true-positive-rate panels implement fixed \emph{cuts on reconstructed quantities} (muon and prong counts, Michel tags, $\pi^{0}$ reconstruction flags, etc.). The global false-positive rate of that cut is computed on the test set; neural models are then evaluated at the same false-positive rate (threshold chosen on the model score) so that bin-wise efficiencies are comparable at matched background rejection.

The cut-based baselines used in the classification plots are:
\begin{itemize}
  \item CC$1\pi^{\pm}$: one reconstructed muon, exactly one charged prong, and exactly one  Michel electron candidate.
  \item CC$N\pi^{\pm}$, $N\geq 1$: one reconstructed muon, at least one charged prong, and at least one Michel candidate.
  \item CC$1\pi^{0}$: one reconstructed muon; a $\pi^{0}$ candidate with two-photon invariant mass ($m_{\gamma \gamma}$) within $\Delta m$ of the $\pi^{0}$ mass ($m_{\pi^{0}} = 134.977\,\mathrm{MeV}$); no Michel electron candidates. We take $\Delta m = m_{\pi^{0}}$, as this maximizes both precision and recall.
\end{itemize}

Charged prongs have PID $8$ (field \verb|prong_part_pid|). The improved Michel electron counting algorithm is used to determine the number of Michel electron candidates (field \verb|improved_nmichel|).

\end{document}